# Computation of magnetization, exchange stiffness, anisotropy, and susceptibilities in large-scale systems using GPU-accelerated atomistic parallel Monte Carlo algorithms

Serban Lepadatu*, George McKenzie, Tim Mercer, Callum Robert MacKinnon, and Philip Raymond Bissell

*Jeremiah Horrocks Institute for Mathematics, Physics and Astronomy, University of Central Lancashire, Preston PR1 2HE, U.K.*

**Abstract**

Monte Carlo algorithms are frequently used in atomistic simulations, including for computation of magnetic parameter temperature dependences in multiscale simulations. Even though parallelization strategies for Monte Carlo simulations of lattice spin models are known, its application to computation of magnetic parameter temperature dependences is lacking in the literature. Here we show how, not only the unconstrained algorithm, but also the constrained atomistic Monte Carlo algorithm, can be parallelized. Compared to the serial algorithms, the parallel Monte Carlo algorithms are typically over 200 times faster, allowing computations in systems with over 10 million atomistic spins on a single GPU with relative ease. Implementation and testing of the algorithms was carried out in large-scale systems, where finite-size effects are reduced, by accurately computing temperature dependences of magnetization, uniaxial and cubic anisotropies, exchange stiffness, and susceptibilities. In particular for the exchange stiffness the Bloch domain wall method was used with a large cross-sectional area, which allows accurate computation of the domain wall width up to the Curie temperature. The exchange stiffness for a simple cubic lattice closely follows an $m^k$ scaling at low temperatures, with $k < 2$ dependent on the anisotropy strength. However, close to the Curie temperature the scaling exponent tends to $k = 2$. Furthermore, the implemented algorithms are applied to the computation of magnetization temperature dependence in granular thin films with over 15 million spins, as a function of average grain size and film thickness. We show the average Curie temperature in such systems may be obtained from a weighted Bloch series fit, which is useful for analysis of experimental results in granular thin films.

* SLepadatu@uclan.ac.uk

# 1. Introduction

Atomistic magnetic simulations are essential for understanding and modelling magnetic processes at the nanoscale, forming a link between *ab-initio* approaches and micromagnetic simulations. In particular computation of temperature dependences of magnetic parameters are used to inform micromagnetic models in a multiscale approach [1-3]. In this respect Monte Carlo algorithms [4,5] are particularly powerful, allowing efficient simulations of thermodynamic properties of spin systems, as well as hysteresis loops [6], and have been implemented in a number of atomistic simulation packages, including Vampire [7], Spirit [8], and Vegas [9]. One current shortcoming, is that the algorithms used are largely serial in nature and cannot be directly implemented on graphical processing units (GPU). GPUs allow massively parallel computations, with processing powers normally available only on distributed computer networks. The serial Monte Carlo algorithms may be used in an embarrassingly parallel approach, or using parallel tempering [10], but this requires a significantly more complicated simulation setup, and the serial algorithms cannot make use of GPUs. Strategies for parallelizing the spin lattice Monte Carlo algorithm on GPUs have been introduced previously, using a red-black checkerboard scheme [11-13], or a stream processing domain decomposition [14]. The application of such parallelization strategies to computation of magnetic parameter temperature dependences is still lacking however, and there is a need for evaluation of their suitability for this important use case. Moreover, one useful variant of the atomistic Monte Carlo algorithm is the constrained Monte Carlo method [15], allowing computation of anisotropy temperature dependence, and a parallelization strategy for this method has not been previously discussed.

Here we implement and test a general-purpose fully parallel (one thread per atomistic spin) adaptive Monte Carlo algorithm, which can be executed both on GPUs and multi-core CPUs. The algorithm was implemented for a simple cubic crystal structure with nearest-neighbour interactions. However, the parallelization scheme can easily be extended to realistic crystal structures and next nearest-neighbour interactions, and we discuss how this can be achieved. Implementation and testing of the algorithm was achieved by computing temperature dependences of magnetic parameters, including magnetization, uniaxial and cubic anisotropies, exchange stiffness, and susceptibilities. Finally, we apply the implemented algorithm to large-scale simulations of the magnetization temperature dependence in granular thin films.

## 2. Parallel Monte Carlo Algorithms

Atomistic spin dynamics uses the Landau-Lifshitz-Gilbert equation [16], supplemented by a stochastic field based on Langevin dynamics [17]. The Fokker-Planck equation corresponding to this stochastic equation has the Boltzmann distribution as a solution [18]. One approach to computing magnetic parameters' temperature dependences, for example as required for micromagnetic models, involves bringing the system into thermodynamic equilibrium, then obtaining statistical information from many stochastically generated atomistic ensembles. In principle it is possible to generate these ensembles using atomistic spin dynamics. However, this approach is very inefficient, particularly close to the phase transition temperature, where a large number of iterations with a very small time step are required to thermalize the system, and generate a large enough set of atomistic ensembles for statistically significant information. A much more efficient approach is to use a Monte Carlo algorithm to generate atomistic ensembles which follow the expected Boltzmann distribution directly; close to phase transition points, which exhibit critical slowing, cluster update algorithms are particularly useful [19,20]. For atomistic simulations a widely used algorithm is based on the Metropolis Monte Carlo algorithm [4]. Here a Monte Carlo iteration, or step, consists in picking each atomistic spin in turn, but in a random order, and rotating it in a small cone about the initial position. This is called a Monte Carlo move, and it is accepted with a probability given by the Boltzmann distribution below.

$$P_i = \exp\left(-\Delta E_i / k_B T\right) \tag{1}$$

The energy change between the previous spin direction and candidate spin move, $\Delta E_i$, is obtained from a spin Hamiltonian, which includes a number of interactions. The energy $E_i$ (J) for spin $i$ with magnetic moment $\mu_S$ and direction $\hat{\mathbf{S}}_i$, is given below.

$$E_i = -\sum_{j \in N_i} J_{ij} \hat{\mathbf{S}}_i . \hat{\mathbf{S}}_j - K_{u,i} \left( \hat{\mathbf{S}}_i . \hat{\mathbf{e}}_i \right)^2 + K_{c,i} \left( \hat{S}_{x,i}^2 \hat{S}_{y,i}^2 + \hat{S}_{x,i}^2 \hat{S}_{z,i}^2 + \hat{S}_{y,i}^2 \hat{S}_{z,i}^2 \right) - \mu_0 \mu_{S,i} \hat{\mathbf{S}}_i . \mathbf{H}_a \tag{2}$$

Here we consider the Heisenberg isotropic exchange interaction with $J_{ij} = J$, where the sum in the first term is over the nearest neighbours of spin $i$. The second and third terms are the uniaxial and cubic anisotropy contributions respectively, where $\hat{\mathbf{e}}_i$ is the easy axis direction, and for cubic anisotropy the easy axes coincide here with the Cartesian axes. The last term is the external field contribution. We have implemented the serial Monte Carlo algorithm in Boris [21] using an adaptive cone angle, similar to the approach used in Ref. [22]. In particular the Monte Carlo move acceptance rate is kept at an optimal value of 0.5 by increasing or decreasing the cone angle in increments of 1 degree as required.

The most straightforward and very efficient approach to parallelizing this algorithm is to use a standard red-black checkerboard ordering scheme [11], which separates the spins into two sets of non-interacting spins, and thus avoids data race conditions if only the nearest neighbours are included in the exchange interaction for a simple cubic structure. However, this scheme can easily be extended to realistic crystal structures, as well as inclusion of next-nearest neighbours, where we simply consider the atomistic spins on non-interacting sub-sets in turn. For example a body-centred cubic (bcc) structure can be considered as 2 interleaved simple cubic sub-lattices, face-centred cubic (fcc) as 4 simple cubic sub-lattices, and hexagonally close packed (hcp) as 4 simple tetragonal sub-lattices. For these realistic crystal structures the red-black ordering scheme can be applied to each sub-lattice in turn, which allows consideration of exchange interactions between any number of spins on different sub-lattices without introducing data race conditions. Thus for fcc and hcp we obtain 8 sub-sets of non-interacting spins, each containing around 1/8$^{\text{th}}$ the total number of atomistic spins. On shared memory machines this results in very efficient parallelization as long as the number of spins in each subset is significantly larger than the number of threads, since the cost of fork-join operations is typically negligible. For GPUs there is an additional cost associated with latency of multiple kernel launches. However, this is also relatively negligible for system sizes containing over 100k atomistic spins. Moreover, since on each red and black subsets the spins are not interacting in Equation (2), we do not need to pick them in a random order, which further reduces the computational cost. We have tested this parallel algorithm by comparison with the serial version, obtaining identical results, and examples are given in the next section. A pseudo-code for this algorithm is given in Appendix A. For systems with 1 million spins and above, the parallel Monte Carlo algorithm is over 200 times faster compared to the serial version, and scales linearly with increasing problem size. The largest problem tested contained 100 million atomistic spins using a 6 GB memory space in single floating-point precision. At this mesh size one Monte Carlo iteration requires ~0.1 s computation time on a RTX 2080 Super GPU, which makes computations feasible on a single GPU even at this extreme system size. In addition, the Monte Carlo algorithm could be implemented for a multi-GPU workstation with an expected near-linear performance scaling, since without the dipole-dipole interaction the communication overhead between multiple GPUs is very small. It is worthy of note that the capability of the latest RTX 3090 GPU means that 1 billion atomistic spins computations are within reach on a single multi-GPU workstation.

A variant of the standard Monte Carlo algorithm is the constrained Monte Carlo algorithm [15], which is useful for computing energies, in particular anisotropy energies, in directions away from an easy axis. This works by generating atomistic ensembles which maintain the average magnetization direction unchanged along the constraining direction, thus effectively working with a sub-set of the possible atomistic ensembles at each thermodynamic equilibrium point. We have also parallelized this algorithm using the red-black ordering scheme, however there are 2 additional difficulties that need to be considered. First, in order to maintain the average magnetization direction unchanged, the algorithm works on pairs of spins, where the change in transverse components is compensated for each spin pair. With the red-black ordering scheme, pairs of spins are picked on the same red or black subsets, which avoids data race conditions. Furthermore, for each Monte Carlo iteration each spin is picked exactly once in a random order, thus forming unique pairs of spins, but different, for each Monte Carlo iteration. We achieve this by shuffling the spin indexes before each iteration, then picking them sequentially in pairs from an array of shuffled indexes. The second difficulty consists in computing the new magnetization length along the constraining direction, since the acceptance probability is proportional to the ratio of the new and old magnetization lengths as shown in Ref. [15]. The difficulty is that once a trial move has been accepted the magnetization length changes. This information could in theory be propagated to all the other computational threads through an atomic operation, however this is very inefficient. Since the change in magnetization length ratio is negligible on average, both at low temperatures and around the phase transition temperature, we resort to simply computing the magnetization length along the constraining direction once at the start of each Monte Carlo iteration. We have tested this parallel algorithm against the serial version as introduced in Ref. [15] with identical results. A pseudo-code for this algorithm is also given in Appendix A.

Finally, it should be noted Equation (2) does not include the long-range dipole-dipole interactions. In many cases this is not necessary, since the exchange energy is typically many orders of magnitude higher. However, in some cases it is important to include this additional interaction, e.g. as used in Refs. [14,23]. This topic is beyond the scope of the current work and will be addressed separately in an upcoming publication.

# 3. Magnetization, Anisotropy, Exchange Stiffness and Susceptibilities

Firstly, we compute the magnetization temperature scaling using the parallel Monte Carlo algorithm, in a cubic mesh with periodic boundary conditions in all directions, thus representative of the bulk magnetization temperature scaling. For the simulations in this work, we use a lattice constant $a$ = 2.5 Å, magnetic moment $\mu_S$ = 1.3 $\mu_B$, exchange energy $J = 6\times10^{-21}$ J, and anisotropy energy $K = 5\times10^{-24}$ J. The result for a 50 $nm^3$ mesh (8 million spins) is shown in Fig. 1(a), where up to 10,000 thermalization steps were used with a temperature step of +1 K close to $T_C$ (increasing temperature sweeps are used), and with averaging over up to 30,000 ensembles for each temperature. The result closely follows the expected Bloch law, Equation **(**3**)**, as shown and from which values of $T_C = 628.0 \pm 0.1$ K, and $\beta = 0.340 \pm 0.001$ are obtained by fitting. The Curie temperature is proportional to the exchange energy, given by $T_C = J\theta_C z/3k_B$, where $z$ is the number of nearest neighbours ($z$ = 6 for a simple cubic lattice), and $\theta_C$ is the reduced Curie temperature. This may be calculated using a self-consistent Gaussian approximation [24], and for a simple cubic lattice with $S = \infty$ Heisenberg model, is given as $\theta_C = 0.723$ [24]. From our computed $T_C$ value, we also obtain $\theta_C = 0.723$ to 3 decimal places, thus in very good agreement. Simulating with a large-scale mesh allows minimization of well-known finite-size effects. Fig. 1(b) shows the computed magnetization scaling as a function of mesh size, from 10$nm^3$ up to 50$nm^3$, illustrating the finite-size effect. Due to the reduced simulation space, a sufficiently thermalized atomistic ensemble cannot be generated at small mesh sizes even with periodic boundary conditions, and the result cannot be improved by increasing the number of ensembles included in the average, or thermalization steps. It should be noted the magnetization temperature dependence in Equation **(**3**)** does not reproduce exactly experimentally measured temperature scaling, particularly at low temperatures, as discussed in Ref. [25]. One approach to accommodating this is to use a temperature rescaling method [26]. In a more recent work, a quantum thermostat was implemented [27], allowing generation of magnon Planck statistics, with results more closely aligned to experimental data compared to Boltzmann statistics; it remains to be investigated how this approach can be combined with a Monte Carlo simulation method.

$$m(T) = \left(1 - T/T_C\right)^{\beta} \quad \textbf{(3)}$$

FIG. 1. (a) Magnetization and anisotropy temperature dependences computed in a 50 nm$^3$ mesh with periodic boundary conditions, using the standard, respectively constrained, parallel Monte Carlo algorithms. The magnetization scaling is fitted using the Bloch law in Equation **(**3**)**, with $T_C$ = 628.0 ± 0.1 K, and Bloch exponent $\beta$ = 0.340 ± 0.001. The solid lines show the Callen-Callen scaling of $m^3$ and $m^{10}$ for uniaxial and cubic anisotropy respectively. The best fit scaling laws have $m$ scaling exponents of 2.95 ± 0.01, and 9.90 ± 0.02 respectively. (b) Detail of $m$ scaling around $T_C$, as a function of simulation mesh size, showing the finite-size effect.

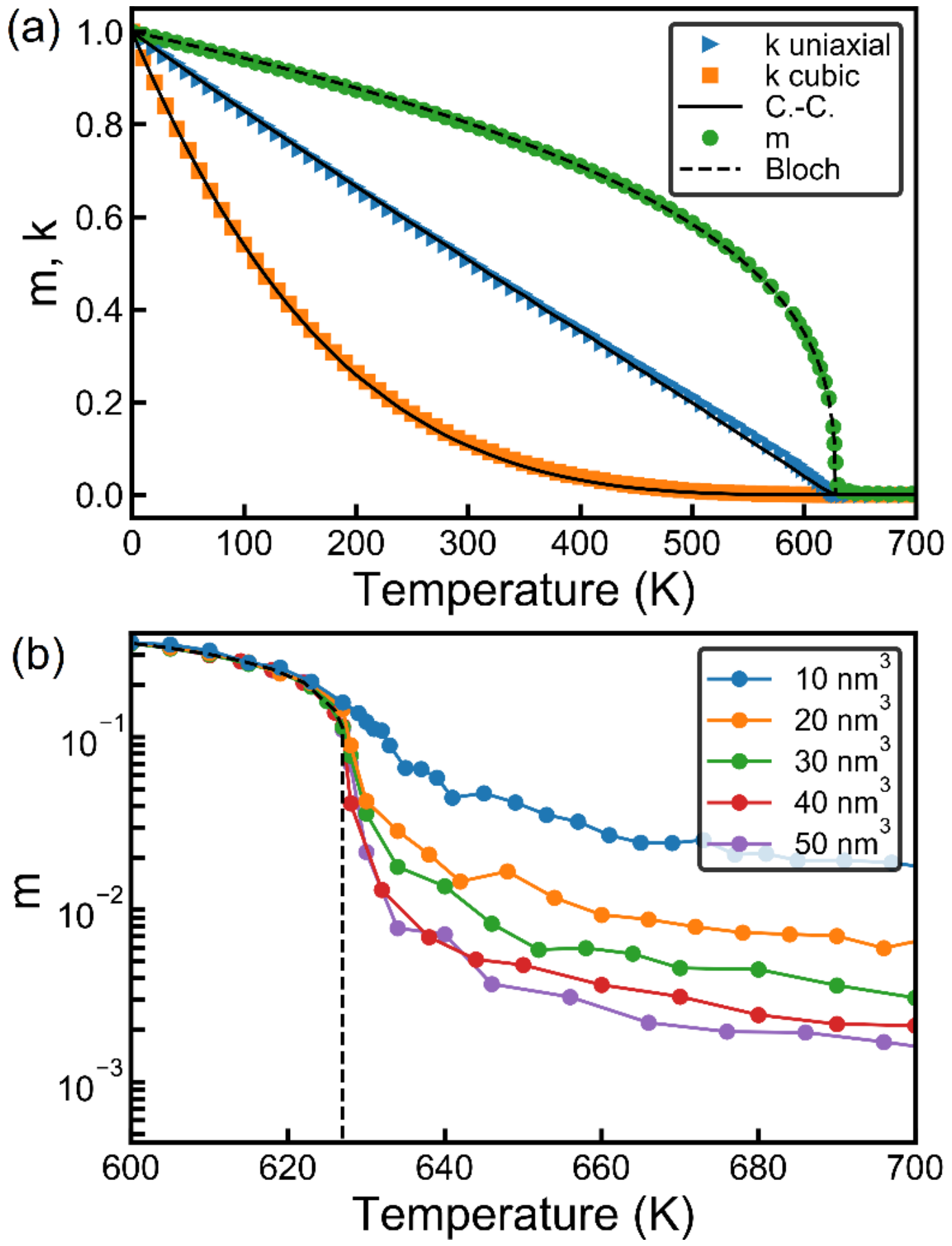

Results for the anisotropy temperature scaling, obtained using the constrained Monte Carlo algorithm, are also shown in Fig. 1(a). The ideal Callen-Callen [28] scaling laws are $m^3$ and $m^{10}$ for uniaxial and cubic anisotropy respectively, and these are shown as solid lines in Fig. 1(a). By fitting the computed $m$ scaling to the computed $k$ scaling we obtain the scaling exponents 2.95 ± 0.01, and 9.90 ± 0.02 respectively. It is well known the computed anisotropy scaling deviates slightly from the ideal Callen-Callen scaling, particularly close to $T_C$, which results in slightly smaller scaling exponents. These results are obtained by computing the average anisotropy energy from Equation **(**2**)**, first along the hard axis (long cube diagonal for the cubic case), then along the easy axis, and finally taking the difference and dividing by the simulation volume to give the un-normalized micromagnetic anisotropy energy density. Another possibility is to compute the torque in a direction away from the easy axis, as done in Ref. [15]. For the cases of uniaxial and cubic anisotropies it is also possible to use the standard Monte Carlo algorithm to compute just the easy axis anisotropy energy, since the hard axis anisotropy is related to the easy axis anisotropy by a simple geometric factor. Thus the scaling laws are obtained using Equation **(**4**)**, $k_u$ for uniaxial anisotropy and $k_c$ for cubic anisotropy, where $K_{0,hard}$ is the zero temperature anisotropy energy density value along the hard axis. We have also verified these cases, obtaining identical results to the constrained Monte Carlo algorithm. The method based on Equation **(**4**)** is faster, since the constrained Monte Carlo algorithm typically requires up to twice longer simulation time due to increased complexity.

$$\begin{aligned} k_u &= 1 - 3K_{easy} / 2K_{0,hard} \\ k_c &= 1 - 5K_{easy} / 3K_{0,hard} \end{aligned} \tag{4}$$

We next discuss the application of the Monte Carlo algorithm to computation of exchange stiffness temperature dependence, $A(T)$. This is a problem of important practical interest, and essential for temperature-dependent micromagnetic models [29]. The problem is complicated by the fact the temperature scaling depends on the crystal structure, thus on the number of neighbours included in the interaction, as well as on the anisotropy strength. Experimental studies on $Ni_2MnIn$ [30] obtained an $m^{1.715}$ scaling at low temperatures, and a much higher scaling of $m^{3.4}$ close to $T_C$. On the other hand experimental results on Fe and Ni [31] obtained a scaling $\sim m^{1.5}$. The exchange stiffness temperature scaling may be computed using the domain wall method discussed in Refs. [32-35]. For a Bloch domain wall with uniaxial anisotropy easy axis along the z-axis, the domain wall width is obtained as shown in Equation **(**5**)**.

$$\begin{aligned} \Delta &= \pi \sqrt{\frac{A(T)}{K(T)}} \\ dF(T) &= 4C\sqrt{A(T)K(T)} \end{aligned} \tag{5}$$

Here Δ is the domain wall width, where the z magnetization component varies along the domain wall as $m \times tanh(-\pi (x-x_0) / \Delta)$. This is shown in Fig. 2 for a 40 $nm^3$ simulation space, where the normalized magnetization z component is obtained for each profile point by averaging over the respective cross-section slice. The computed profiles closely follow the expected *tanh* function, even for $T/T_C = 0.99$. We note that for temperatures close to $T_C$ a large cross-section area is required to obtain a sufficiently averaged profile. In Equation **(**5**)** $C$ is the cross-section area perpendicular to the domain wall, and $dF(T)$ is the change in free energy between the state with a domain wall and the uniform state, obtained from the change in internal energy – Equation **(**2**)** – as given in Ref. [32]. In Equation **(**5**)** $K(T)$ may be eliminated, thus obtaining $A(T)$ from the fitted values of $\Delta(T)$ and the computed $dF(T)$. However, we find the $dF(T)$ values tend to be inaccurate, particularly at low temperatures, since they rely on taking the difference between large energy values, and numerically integrating with increasingly large steps as $T \rightarrow 0$. Instead, we first accurately compute $K(T)$ using a uniform state temperature sweep as detailed above, thus obtaining $A(T)$ from the fitted $\Delta(T)$ values. Above $T_C$ however, where a $\Delta$ value is not available, we compute the exchange stiffness only from the $dF(T)$ values, using the $K(T)$ values obtained separately.

FIG. 2. Domain wall profiles showing the $m_z$ component at different temperatures, together with a *tanh* fit used to obtain the domain wall width. The inset shows a Bloch domain wall in a 40 nm$^3$ computational mesh, with green arrows pointing up, and purple arrows pointing down.

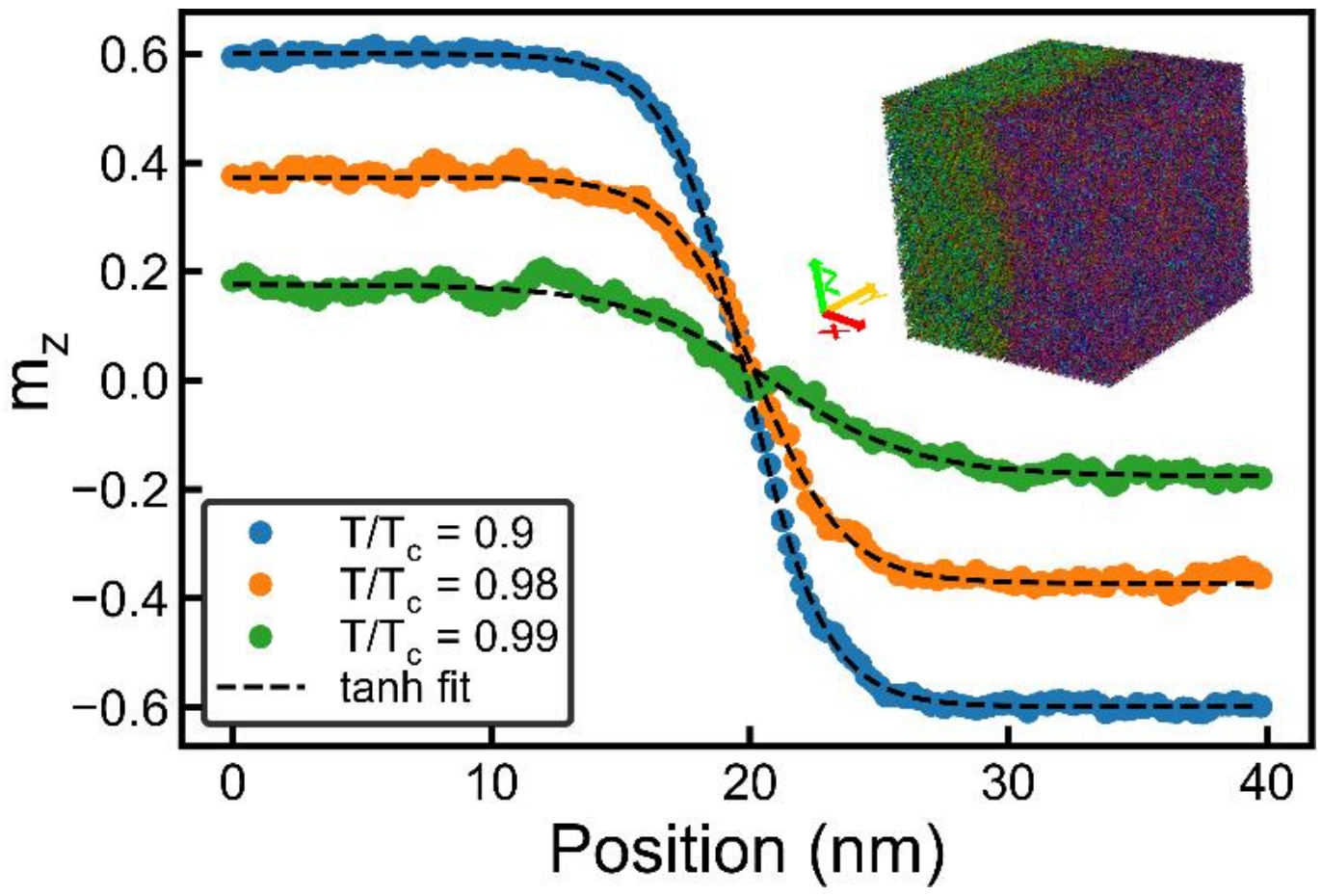

Results are shown in Fig. 3(a) for a 40 nm$^3$ mesh, with an anisotropy energy value of $K = 5\times10^{-23}$ J. For these simulations periodic boundary conditions were used along y and z directions, and anti-periodic along x direction for the domain wall temperature sweep. The best fit to the low-temperature exchange stiffness scaling ($m > 0.6$) is an $m^{1.48}$ dependence, and as can be observed the computed temperature scaling closely follows the $m^k$ scaling law except when approaching $T_C$. In the mean field approximation (MFA) $A$ scales as $m^2$. However, due to spin-spin correlations, in general the scaling exponent is reduced as $m^{2-\varepsilon}$, a result which is reproduced by the classical spectral density method [34]. As the temperature tends to $T_C$, the MFA result is recovered however, as the effect of spin-spin correlations is reduced. To see this clearly we plot $A$ as a function of $m$ in Fig. 3(b), indicating both the $m^2$ and $m^{1.48}$ scalings for comparison. As can be seen, as the temperature tends towards $T_C$ there is a rapid transition from the $m^{1.48}$ low-temperature scaling to the $m^2$ scaling. We note the exponent value of 1.48 is in good agreement with experimental results on Fe and Ni [31].

FIG. 3. (a) Exchange stiffness temperature dependence, shown for $K = 5\times10^{-23}$ J, computed using a 40 nm$^3$ mesh with the Bloch domain wall method, and with fitted domain wall width averaged over 50,000 Monte Carlo steps at each temperature point. The best fit low-temperature $m$ scaling law with exponent 1.48 ± 0.02 is shown. However, close to the Curie temperature a discrepancy can be observed. (b) Un-normalized exchange stiffness plotted as a function of $m$, showing the temperature scaling tends to $m^2$ when approaching the Curie temperature.

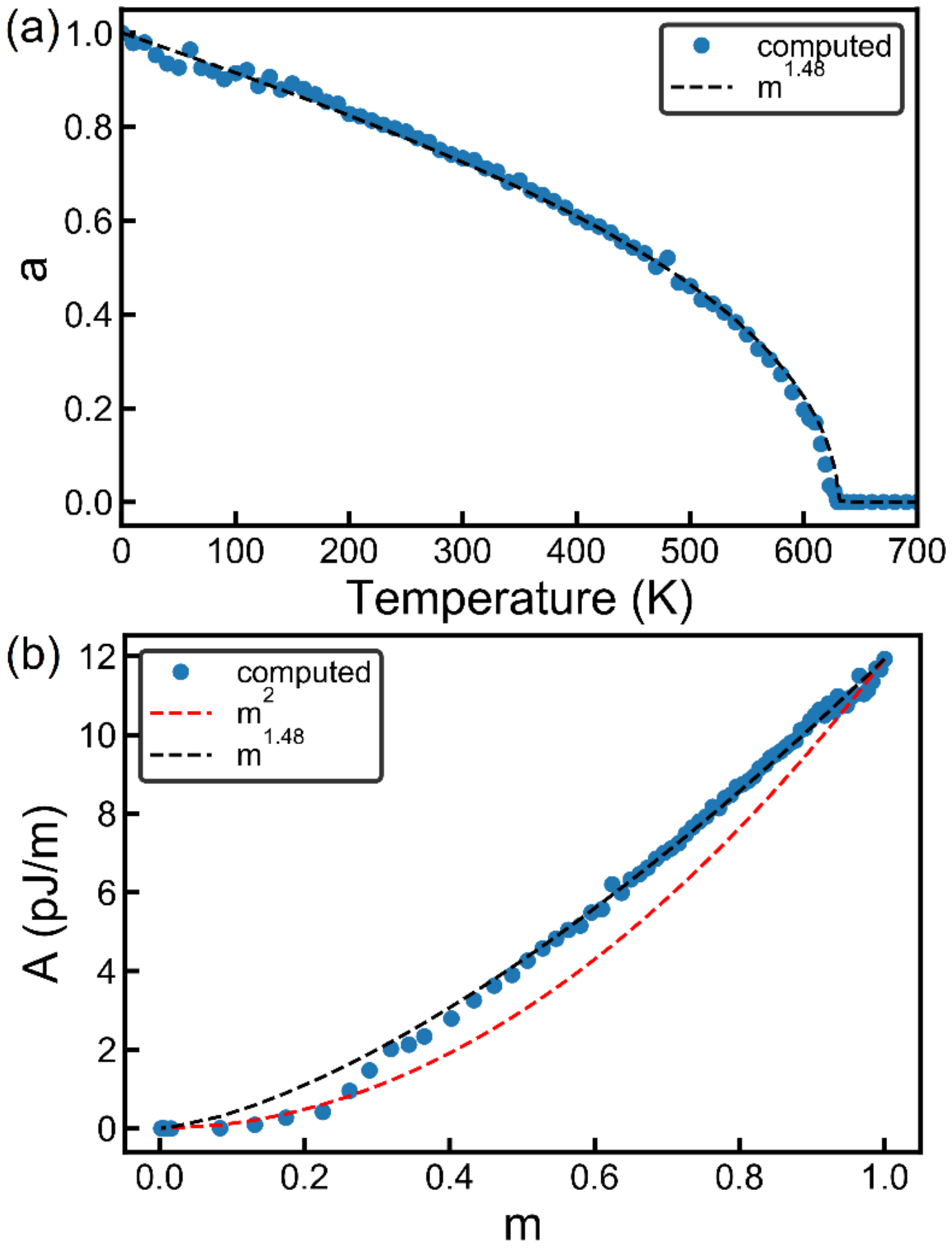

In general, the temperature dependence of the exchange stiffness is a complex problem, and is strongly material-dependent, differing depending on the crystal structure as well as anisotropy strength. Thus, whilst an $m^k$ scaling is observed, for high-anisotropy materials such as FePt, $k$ = 1.76 was obtained, whilst for a simple cubic structure with high anisotropy $k$ = 1.66 was obtained in the low-temperature range [34]. On the other hand, a smaller value of $k$ = 1.54 was obtained for a cubic monolayer [36], whilst a value of $k$ = 1.55 was obtained for $Nd_2Fe_{14}B$ [37], decreasing to $k$ = 1.2 at low temperatures. To investigate this problem further, we repeated the computations by varying the anisotropy strength. As the anisotropy strength increases, the low-temperature scaling exponent also increases, as the effect of spin-spin correlations is reduced. The results are summarised in Table I, noting a good agreement is obtained with the value of $k$ = 1.66 obtained for a simple cubic structure with large anisotropy in Ref. [34]. The results shown here were obtained using the Monte Carlo algorithm to thermalize the system and obtain an average Δ value. However, the same results are obtained when using the atomistic spin dynamics approach, although this requires significantly longer computation time particularly close to $T_C$. The computations were performed in single floating-point precision on the GPU. However, we have also verified the same results are obtained when running in double floating-point precision.

TABLE I. Computed low-temperature $m^k$ scaling of exchange stiffness as a function of anisotropy strength for a simple cubic structure.

| $K/J$ | $8.33\times10^{-3}$ | $1.67\times10^{-2}$ | $2.5\times10^{-2}$ | $3.33\times10^{-2}$ | $4.2\times10^{-2}$ |
|---|---|---|---|---|---|
| $k$ | 1.48 ± 0.02 | 1.55 ± 0.02 | 1.75 ± 0.02 | 1.75 ± 0.02 | 1.79 ± 0.02 |

Finally, we test the Monte Carlo algorithm by computing the temperature dependence of susceptibilities. This is obtained by computing the variance of the average normalized magnetization in a large set of atomistic ensembles as shown in Equation (6).

$$\tilde{\chi}_\alpha(T) = \frac{N\mu_S}{k_B T}\left[\left\langle \overline{m}_\alpha^2 \right\rangle - \left\langle \overline{m}_\alpha \right\rangle^2\right], \quad \alpha = l, x, y, z \tag{6}$$

Here $\overline{m}_\alpha$ is the normalized magnetization, obtained by a spatial average in a single atomistic ensemble, and the operation denoted by angular brackets is averaging over many atomistic ensembles. $N$ is the number of atomistic spins in the computational mesh, and $\alpha = l, x, y, z$ denotes the magnetization component, such that $m_l \equiv m$ is the magnetization length, and $\tilde{\chi}_l(T)$ is the relative longitudinal susceptibility (units 1/Tesla). Results for $K = 5\times10^{-23}$ J in a 30nm$^3$ mesh, are shown in Fig. 4. The longitudinal susceptibility follows the expected temperature dependence, tending to zero as $T$ tends to 0 K, and showing a critical behaviour at $T_C$, although for a system with finite number of spins $N$, the susceptibilities are always finite. Whilst the easy axis susceptibility matches the longitudinal susceptibility below $T_C$, above $T_C$ both the hard-axis and easy-axis susceptibilities are the same as expected. The hard-axis (or transverse) susceptibility is plotted in Fig. 4 as the average of $\tilde{\chi}_y(T)$ and $\tilde{\chi}_z(T)$, noting the easy axis is along the x axis. This is compared to the temperature dependence obtained from the computed $m$ and $K_u(T)$, based on Equation **(7)** for $T < T_C$, showing a good agreement. Note for the atomistic parameters used, we have $K_u(0) = 3.2$ MJ/m$^3$, and $M_S(0) = 771598$ A/m.

FIG. 4. Susceptibilities computed using a 30 nm$^3$ mesh and uniaxial anisotropy along the x axis, with K = $5\times10^{-23}$ J. The variance in Equation **(6)** is obtained from a set of 50,000 atomistic ensembles at each temperature. For the longitudinal susceptibility $\alpha = l$, for the easy axis susceptibility $\alpha = x$, and for the hard axis susceptibility (transverse susceptibility) we average the results for $\alpha = y$ and $\alpha = z$. The latter is compared to the transverse susceptibility obtained from Equation **(7)**, shown as a solid line, with $m(T)$ and $K_u(T)$ computed separately for the same mesh size. The vertical dashed line indicates the Curie temperature of 628.0 K.

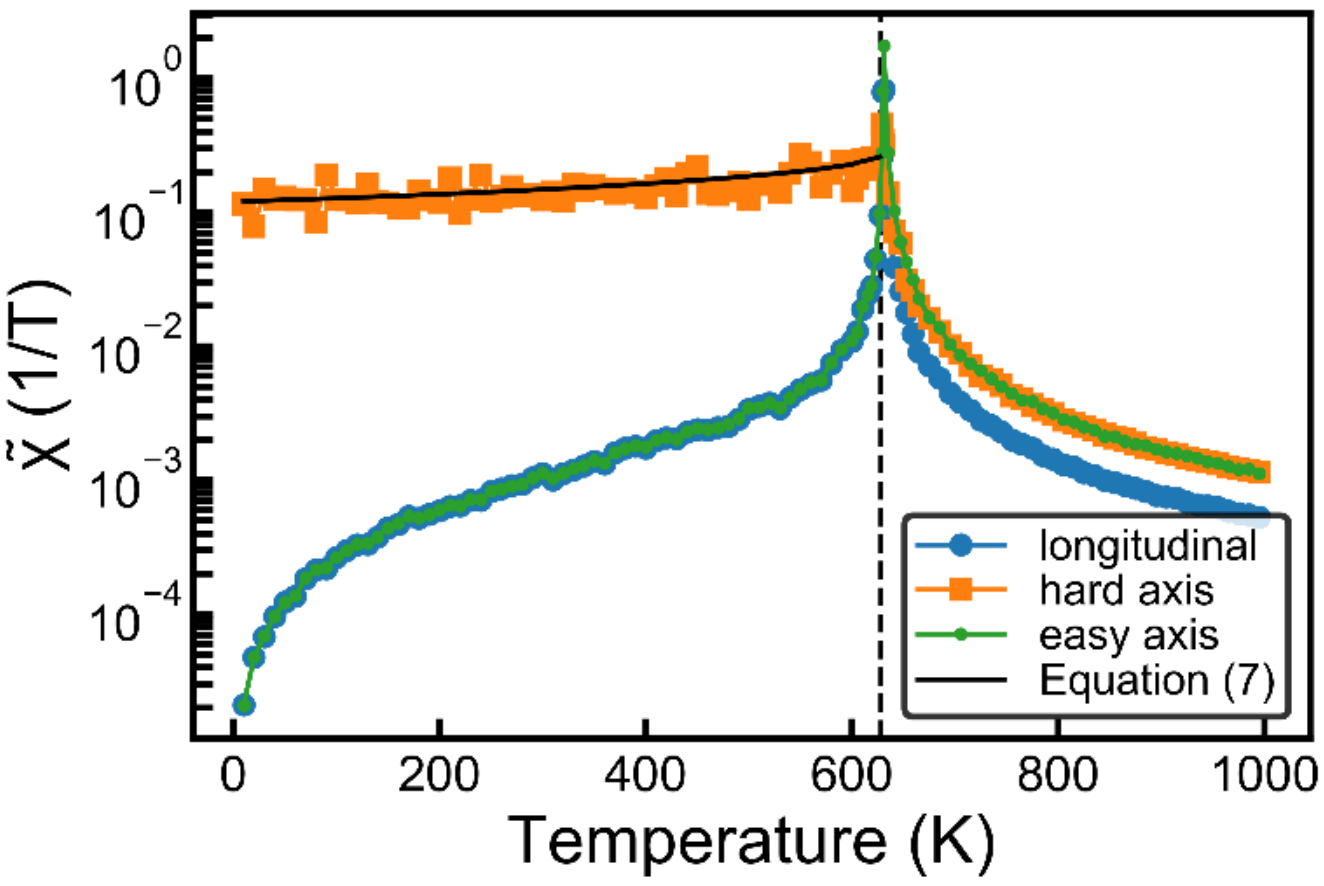

$$\tilde{\chi}_{\perp,u}(T) = \frac{M_S(0)m^2(T)}{2K_u(T)}, \quad T < T_C \tag{7}$$

## 4. Large-Scale Granular Thin Films

The implemented parallel Monte Carlo algorithm is now applied to a large-scale problem, in particular calculation of magnetization temperature dependence in granular thin films, in order to illustrate the need for large-scale Monte Carlo simulations. Previous works studied the grain size influence on magnetic behaviour in polycrystalline thin films using both field cooled and zero-field cooled atomistic simulations [38,39]. Here we concentrate on simulating the magnetization temperature dependence using an increasing temperature sweep, in order to analyse the phase transition temperature in granular thin films. This is useful for example in heat-assisted magnetic recording [40] and magnetic nanoparticles [41]. A previous study has shown how a $T_C$ distribution in granular materials may be extracted using an integral method [42]. An alternative approach is to fit the magnetization temperature dependence using a weighted Bloch series, which allows definition of an average Curie temperature.

FIG. 5. Illustration of simulated granular thin film with 200 nm$^2$ in-plane area and average 20 nm$^2$ grains (20 nm grain size), with the grains obtained using Voronoi tessellation. The image shows the y components of atomistic moments at room temperature, with red indicating +y direction and blue indicating –y direction.

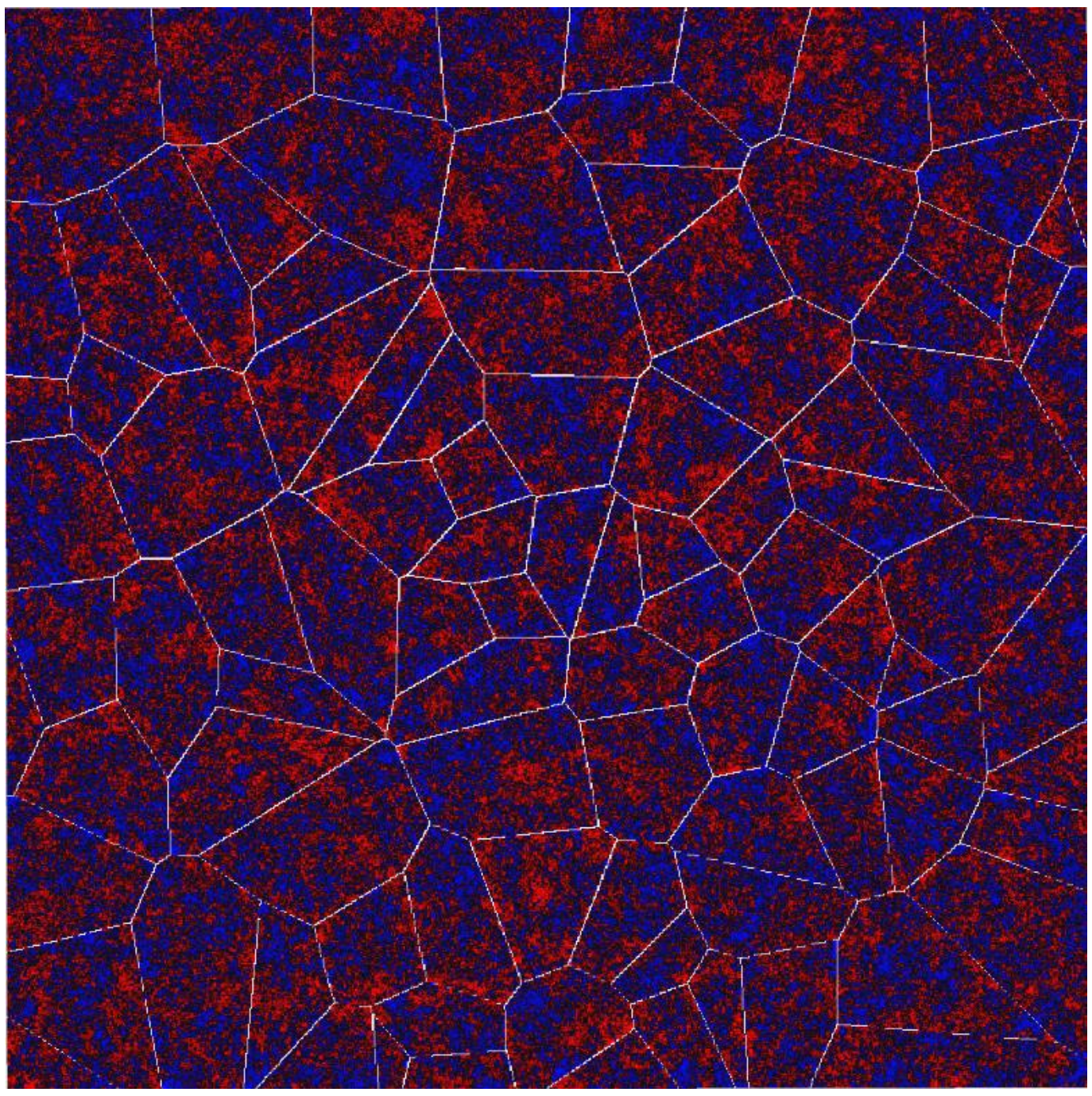

For these simulations we used a 200 nm$^2$ in-plane area with in-plane periodic boundary conditions, and thickness values ranging from 1 nm to 6 nm, without periodic boundary conditions perpendicular to the film. A grain structure was generated using Voronoi tessellation, with average grain size varying from 5 nm to 40 nm – an example for a 20 nm average grain size is shown in Fig. 5. Since the Curie temperature is strongly dependent on the grain size, the magnetization in such thin films does not follow a simple Bloch law. Results for a 2 nm thick film are shown in Fig. 6 as a function of average grain size.

FIG. 6. Computed magnetization scaling in a 2 nm granular thin film, as a function of average grain size indicated in the legend. The magnetization scaling is fitted using the multi-Bloch scaling in Equation **(**8**)**, indicated as dashed lines.

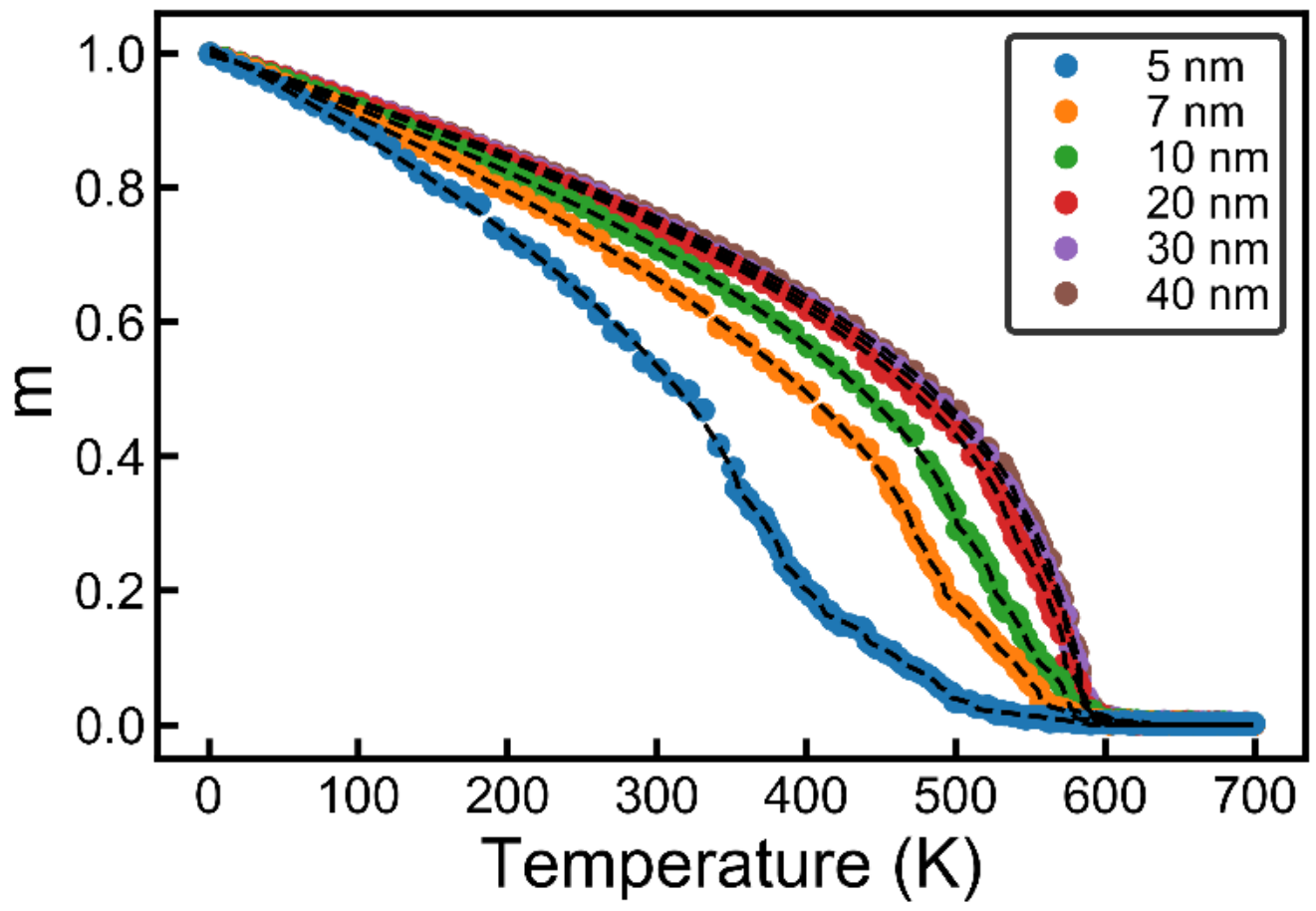

Whilst individual grains follow a Bloch law, the entire system can be described by a weighted series of Bloch laws, as given in Equation **(**8**)**. Here each grain contributes a different Bloch term in the sum, with weight $w_i$ proportional to its area, noting the sum of the weights equals 1. Moreover, since many grains have approximately the same size, significantly fewer Bloch terms, $N_B$, are required in Equation **(**8**)** than the number of grains. The computed magnetization temperature dependence in Fig. 6 is fitted using Equation **(**8**)**, with fits shown as dashed lines. The fits are performed as a function of increasing number of Bloch terms $N_B$, and the results converge for $N_B > 10$, with the obtained Bloch terms $T_{C,i}$ values being smaller than the corresponding continuous thin film values, as expected. This allows an average $\overline{T}_C$ value to be obtained, given in Equation **(**8**)** as a weighted average of the individual Bloch term values.

$$m(T) = \sum_{i=1}^{N_B} w_i \left(1 - T / T_{C,i}\right)^{\beta,i}$$
$$\overline{T}_C \equiv \sum_{i=1}^{N_B} w_i T_{C,i} \tag{8}$$

This analysis was performed as a function of film thickness and average grain size, with results shown in Fig. 7. Here the dashed horizontal lines represent the continuous thin film $T_C$ values at each thickness. As expected, increasing the average grain size results in increasing average $\overline{T}_C$ values, tending towards the continuous thin film value in the limit of large grain size. Moreover, increasing the film thickness results in increasing $\overline{T}_C$ values, tending towards the bulk value of 628.0 K computed in the previous section, in the limit of thick films and large grain size. These results show how an experimentally obtained magnetization temperature dependence for granular thin films may be analysed in order to obtain an average Curie temperature value.

FIG. 7. Average $T_C$, obtained using Equation **(**8**)**, computed as a function of average grain size and film thickness. The horizontal dashed lines indicated the $T_C$ values for the continuous thin films at each respective thickness.

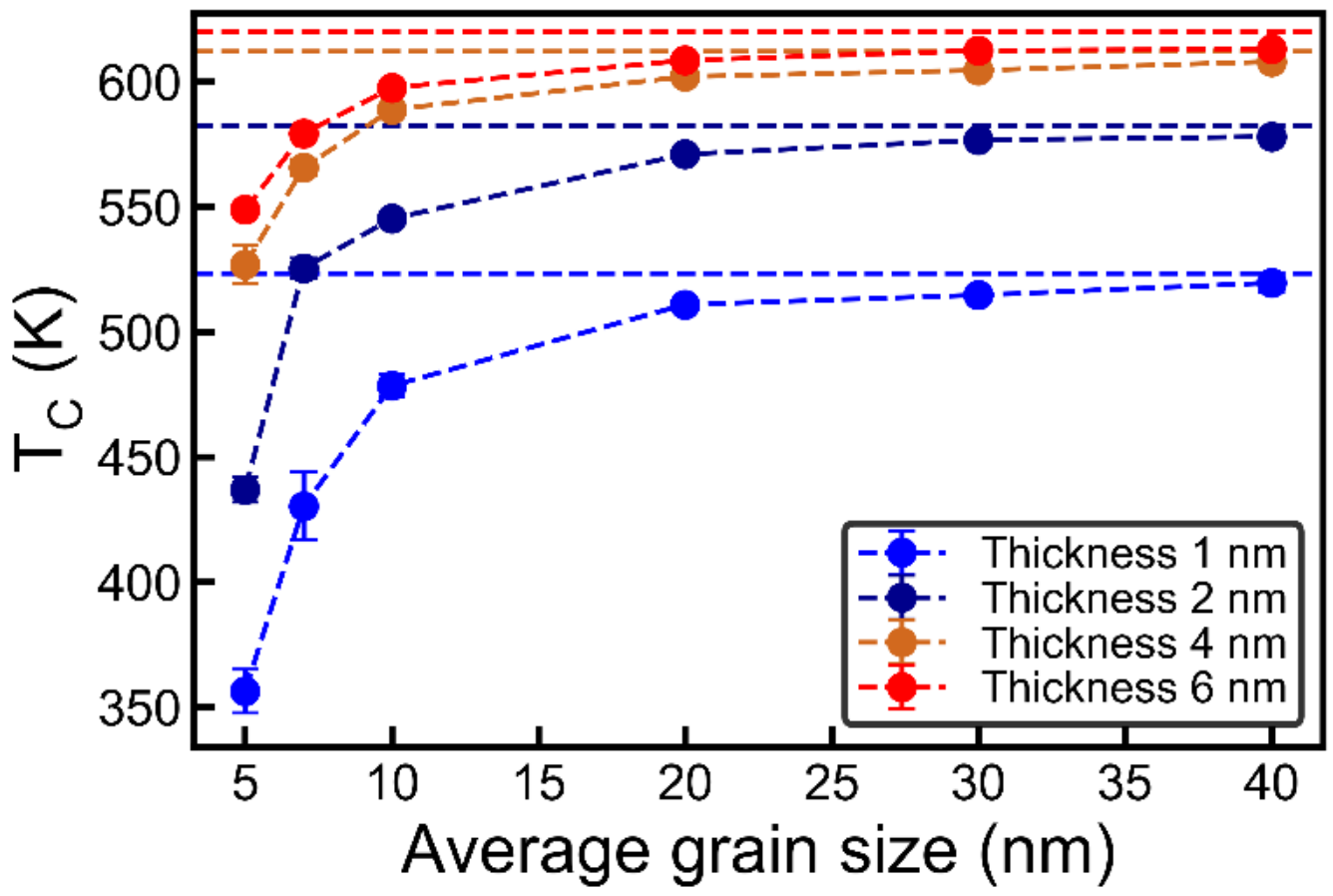

## 5. Conclusions

In this work we implemented and tested a general-purpose fully parallel atomistic Monte Carlo algorithm, based on the Metropolis Monte Carlo algorithm, as well as the previously introduced constrained Monte Carlo algorithm. Since our approach allows for one thread per atomistic spin – or spin pair in the case of the constrained variant – the algorithms can be implemented efficiently on GPUs, and for systems containing over 1 million spins we obtained speedup factors of over 200 compared to the serial versions. Whilst the algorithms implemented here were tested for a simple cubic crystal structure, the parallelization approach can easily be extended to realistic crystal structures. The algorithms were tested by computing temperature dependences of magnetic parameters in large-scale systems where finite-size effects are reduced, including magnetization, uniaxial and cubic anisotropy, exchange stiffness, and susceptibilities. In particular for the exchange stiffness, using a large cross-sectional area allows accurate computation of the domain wall width up to $T_C$. In a simple cubic structure the exchange stiffness follows an $m^k$ scaling law at low temperatures, with $k < 2$. As the temperature tends to $T_C$, the mean field approximation result of $k = 2$ is recovered. Moreover, the $k$ exponent has a marked dependence on the anisotropy strength, increasing as the anisotropy strength increases. The implemented Monte Carlo algorithm was also applied to computation of magnetization temperature dependence in large-scale granular thin films as a function of average grain size and film thickness. Using a weighted Bloch series fit we demonstrated how a Curie temperature distribution for the individual grains may be extracted, allowing a simple definition of the average Curie temperature. This method may readily be applied to experimentally measured magnetization temperature dependences in granular thin films.

# Appendix A: Parallel Monte Carlo Algorithms Pseudocode

Pseudocode for the parallel Monte Carlo algorithm is shown in Table A1, and for the constrained parallel Monte Carlo algorithm in Table A2.

TABLE A1. Parallel Monte Carlo algorithm. Here UniformRandom([0, 1]) generates a random number in the interval [0, 1] with uniform probability distribution, and UniformRandom_RotateSpin($\mathbf{S}_i$, θ) generates a rotated spin about spin $\mathbf{S}_i$, within a cone with angle θ, with uniform probability distribution.

```
DATA:
Cone angle: θ
Temperature: T
Number of spins N = N_red + N_black
Atomistic spins: S_i := S_1, …, S_N

procedure MonteCarlo_Red(Black)ParallelPass(S_1, …, S_N, T, θ)
        for index in range 1, …, N_red(black) do parallel
                i = Get_Red(Black)SpinIndex(index)
                S_new,i = UniformRandom_RotateSpin(S_i, θ)
                P = UniformRandom([0, 1])
                if P < exp[-(E(S_new,i) – E(S_i)) / k_B T] then
                        S_i = S_new,i
                        reduction: acceptance_rate += 1 / N
                end if
        end for
        return acceptance_rate
end procedure

MONTE CARLO ALGORITHM:
acceptance_rate = MonteCarlo_RedParallelPass(S_1, …, S_N, T, θ)
acceptance_rate += MonteCarlo_BlackParallelPass(S_1, …, S_N, T, θ)
if acceptance_rate > 0.5 then
   if θ < 180 then θ += 1 end if
else
   if θ > 1 then θ -= 1 end if
end if
```

TABLE A2. Parallel constrained Monte Carlo algorithm. Here RotateCoordinateSystem($\tilde{\mathbf{S}}_i$, $\tilde{\mathbf{S}}_{i+1}$, $\mathbf{e}_C$) generates rotated spins from $\tilde{\mathbf{S}}_i$, $\tilde{\mathbf{S}}_{i+1}$, such that the new spin components are represented in a coordinate system with x-axis in the direction $\mathbf{e}_C$ in the original coordinate system. InverseRotateCoordinateSystem is the inverse coordinate system rotation operation. ShuffleRedBlackSpins shuffles the spins in the red and black checkerboards.

**DATA**:
Cone angle: $\theta$
Constraining Direction: $\mathbf{e}_C$
Number of spins: $N = N_{red} + N_{black}$
Atomistic spins: $\mathbf{S}_i := \mathbf{S}_1, \ldots, \mathbf{S}_N$
Atomistic spins on red/black checkerboard in shuffled order: $\tilde{\mathbf{S}}_{r1}, \ldots, \tilde{\mathbf{S}}_{rNred}, \tilde{\mathbf{S}}_{b1}, \ldots, \tilde{\mathbf{S}}_{bNblack}$

---

**procedure** ConstrainedMonteCarlo_ParallelPass($\tilde{\mathbf{S}}_1, \ldots, \tilde{\mathbf{S}}_{\tilde{N}}$, T, $\theta$, $M_L$, $\mathbf{e}_C$)
    **for** i in range 1, …, $\tilde{N}$ **step** 2 **do parallel**
        $\hat{\mathbf{S}}_i, \hat{\mathbf{S}}_{i+1}$ = RotateCoordinateSystem($\tilde{\mathbf{S}}_i, \tilde{\mathbf{S}}_{i+1}, \mathbf{e}_C$)
        $\hat{\mathbf{S}}_{new,i}$ = UniformRandom_RotateSpin($\hat{\mathbf{S}}_i$, $\theta$)
        $^{(y)}\hat{S}_{new,i+1} = {}^{(y)}\hat{S}_{i+1} + {}^{(y)}\hat{S}_i - {}^{(y)}\hat{S}_{new,i}$
        $^{(z)}\hat{S}_{new,i+1} = {}^{(z)}\hat{S}_{i+1} + {}^{(z)}\hat{S}_i - {}^{(z)}\hat{S}_{new,i}$
        $^{(x)}\hat{S}_{new,i+1} = \mathrm{sign}({}^{(x)}\hat{S}_{i+1})[\hat{\mathbf{S}}_{i+1}^2 - {}^{(y)}\hat{S}_{new,i+1}^2 - {}^{(z)}\hat{S}_{new,i+1}^2]^{0.5}$
        $\hat{M}_L = |M_L| + {}^{(x)}\hat{S}_{new,i} + {}^{(x)}\hat{S}_{new,i+1} - {}^{(x)}\hat{S}_i - {}^{(x)}\hat{S}_{i+1}$
        $\tilde{\mathbf{S}}_{new,i}, \tilde{\mathbf{S}}_{new,i+1}$ = InverseRotateCoordinateSystem($\hat{\mathbf{S}}_{new,i}, \hat{\mathbf{S}}_{new,i+1}, \mathbf{e}_C$)
        P = UniformRandom([0, 1])
        **if** $P < (\hat{M}_L / M_L)^2|{}^{(x)}\hat{S}_{i+1} / {}^{(x)}\hat{S}_{new,i+1}|\exp[-(E(\tilde{\mathbf{S}}_{new,i})+E(\tilde{\mathbf{S}}_{new,i+1})-E(\tilde{\mathbf{S}}_i)-E(\tilde{\mathbf{S}}_{i+1})/k_BT]$ **then**
            $\tilde{\mathbf{S}}_i, \tilde{\mathbf{S}}_{i+1} = \tilde{\mathbf{S}}_{new,i}, \tilde{\mathbf{S}}_{new,i+1}$
            **reduction:** acceptance_rate += 2 / N
        **end if**
    **end for**
    **return** acceptance_rate
**end procedure**

---

**CONSTRAINED MONTE CARLO ALGORITHM**:
$M_L = \mathrm{Sum}_{i=1,\ldots,N}(\mathbf{S}_i.\mathbf{e}_C)$
$\tilde{S}_{r1}, \ldots, \tilde{S}_{rNred}, \tilde{S}_{b1}, \ldots, \tilde{S}_{bNblack}$ = ShuffleRedBlackSpins($S_1, \ldots, S_{Nred}, S_1, \ldots, S_{Nblack}$)
acceptance_rate = ConstrainedMonteCarlo_ParallelPass($\tilde{S}_{r1}, \ldots, \tilde{S}_{rNred}$, T, $\theta$, $M_L$, $\mathbf{e}_C$)
acceptance_rate += ConstrainedMonteCarlo_ParallelPass($\tilde{S}_{b1}, \ldots, \tilde{S}_{bNblack}$, T, $\theta$, $M_L$, $\mathbf{e}_C$)
**if** acceptance_rate > 0.5 **then**
   **if** $\theta$ < 180 **then** $\theta$ += 1 **end if**
**else**
   **if** $\theta$ > 1 **then** $\theta$ -= 1 **end if**
**end if**

# References

[1] R.E. Rottmayer et al., “Heat-assisted magnetic recording” IEEE Trans. Magn. **42**, 2417 (2006).

[2] S.C. Westmoreland et al., “Multiscale model approaches to the design of advanced permanent magnets” Scripta Materialia **148**, 56 (2018).

[3] A. Meo, W. Pantasri, W. Daeng-am, S.E. Rannala, S.I. Ruta, R.W. Chantrell, P. Chureemart, and J. Chureemart, “Magnetization dynamics of granular heat-assisted magnetic recording media by means of a multiscale model” Phys. Rev. B **102**, 174419 (2020).

[4] N. Metropolis, A.W. Rosenbluth, M.N. Rosenbluth, A.H. Teller, and E. Teller, “Equation of State Calculations by Fast Computing Machines” J. Chem. Phys. **21**, 1087 (1953).

[5] D. Hinzke and U. Nowak, “Magnetization switching in a Heisenberg model for small ferromagnetic particles” Phys. Rev. B **58**, 265 (1998).

[6] Z. Nehme, Y. Labaye, R. Sayed Hassan, N. Yaacoub, and J.M. Greneche, “Modeling of hysteresis loops by Monte Carlo simulation” AIP Advances **5**, 127124 (2015).

[7] R.F.L. Evans, W.J. Fan, P. Chureemart, T.A. Ostler, M.O.A. Ellis, and R.W. Chantrell, “Atomistic spin model simulations of magnetic nanomaterials” J. Phys.: Condens. Matter **26**, 103202 (2014).

[8] G.P. Müller, M. Hoffmann, C. Dißelkamp, D. Schürhoff, S. Mavros, M. Sallermann, N.S. Kiselev, H. Jónsson, and S. Blügel, “Spirit: Multifunctional framework for atomistic spin simulations” Phys. Rev. B **99**, 224414 (2019).

[9] J.D. Alzate-Cardona, D. Sabogal-Suarez, O.D. Arbelaez-Echeverri, and E. Restrepo-Parra, “VEGAS: Software package for the atomistic simulation of magnetic materials” Revista Mexicana de Fısica **64**, 490 (2018).

[10] D.J. Earl and M.W. Deema, “Parallel tempering: Theory, applications, and new perspectives” Phys. Chem. Chem. Phys. **7**, 3910 (2005).

[11] T. Preis, P. Virnau, W. Paul, and J.J. Schneider, “GPU accelerated Monte Carlo simulation of the 2D and 3D Ising model” *Journal of Computational Physics* **228**, 4468-4477 (2009).

[12] B. Block, P. Vinau, and T. Preis, “Multi-GPU accelerated multi-spin Monte Carlo simulations of the 2D Ising model” *Computer Physics Communications* **181**, 1549-1556 (2010).

[13] M. Weigel, T. Yavors'kii, "GPU accelerated Monte Carlo simulations of lattice spin models" *Physics Procedia* **15**, 92-96 (2011).
[14] M.C. Ambrose and R.L. Stamps, "Monte Carlo simulation of the effects of higher-order anisotropy on the spin reorientation transition in the two-dimensional Heisenberg model with long-range interactions" *Phys. Rev. B* **87**, 184417 (2013).
[15] P. Asselin, R.F.L. Evans, J. Barker, R.W. Chantrell, R. Yanes, O. Chubykalo-Fesenko, D. Hinzke, and U. Nowak, "Constrained Monte Carlo method and calculation of the temperature dependence of magnetic anisotropy" Phys. Rev. B **82**, 054415 (2010).
[16] T.L. Gilbert, "A Lagrangian formulation of the gyromagnetic equation of the magnetic field" Phys. Rev. **100**, 1243 (1955).
[17] W.F. Brown Jr., "Thermal fluctuation of fine ferromagnetic particles" IEEE Trans. Magn. **15**, 1196 (1979).
[18] D.A. Garanin, "Fokker-Planck and Landau-Lifshitz-Bloch equations for classical ferromagnets" Phys. Rev. B **55**, 3050 (1997).
[19] R.H. Swendsen and J.-S. Wang, "Nonuniversal critical dynamics in Monte Carlo simulations" *Phys. Rev. Lett.* **58**, 86 (1987).
[20] U. Wolff, "Collective Monte Carlo updating for spin systems" *Phys. Rev. Lett.* **62**, 361 (1989).
[21] S. Lepadatu, "Boris computational spintronics - High performance multi-mesh magnetic and spin transport modeling software" J. Appl. Phys. **128**, 243902 (2020).
[22] J.D. Alzate-Cardona, D. Sabogal-Suárez, R.F.L. Evans, and E. Restrepo-Parra, "Optimal phase space sampling for Monte Carlo simulations of Heisenberg spin systems" J. Phys.: Condens. Matter **31**, 095802 (2019).
[23] D.A. Wahab et al., "Quantum Rescaling, Domain Metastability, and Hybrid Domain-Walls in 2D CrI3 Magnets" Advanced Materials **20**, 04138 (2020).
[24] D.A. Garanin, "Self-consistent Gaussian approximation for classical spin systems: Thermodynamics" Phys. Rev. B **53**, 593 (1996).
[25] M.D. Kuz'min, "Shape of Temperature Dependence of Spontaneous Magnetization of Ferromagnets:Quantitative Analysis" Phys. Rev. Lett. **94**, 107204 (2005).
[26] R.F.L. Evans, U. Atxitia, and R.W. Chantrell, "Quantitative simulation of temperature-dependent magnetization dynamics and equilibriumproperties of elemental ferromagnets" Phys. Rev. B **91**, 144425 (2015).
[27] J. Barker and G.E.W. Bauer, "Semiquantum thermodynamics of complex ferrimagnets" Phys. Rev. B **100**, 14040 (2019).

[28] H.B. Callen and E. Callen, “The present status of the temperature dependence of magnetocrystalline anisotropy, and the l(l+1)2 power law” J. Phys. Chem. Solids **27**, 1271 (1966).
[29] S. Lepadatu, “Emergence of transient domain wall skyrmions after ultrafast demagnetization” Phys. Rev. B **102**, 094402 (2020).
[30] K. Niitsu, X. Xu, R.Y. Umetsu, R. Kainuma, and K. Harada, “Temperature dependence of exchange stiffness in an off-stoichiometric Ni2MnIn Heusler alloy” Phys. Rev. B **101**, 014443 (2020).
[31] K. Niitsu, “Temperature dependence of magnetic exchange stiffness in iron and nickel” J. Phys. D: Appl. Phys. **53**, 39LT01 (2020).
[32] N. Kazantseva, R. Wieser, and U. Nowak, “Transition to Linear Domain Walls in Nanoconstrictions” Phys. Rev. Lett. **94**, 037206 (2005).
[33] D. Hinzke, N. Kazantseva, U. Nowak, O.N. Mryasov, P. Asselin, and R.W. Chantrell, “Domain wall properties of FePt: From Bloch to linear walls” Phys. Rev. B **77**, 094407 (2008).
[34] U. Atxitia et al. “Multiscale modeling of magnetic materials: Temperature dependence of the exchange stiffness” Phys. Rev. B **82**, 134440 (2010).
[35] R. Moreno, R.F.L. Evans, S. Khmelevskyi, M.C. Munoz, R.W. Chantrell, and O. Chubykalo-Fesenko, “Temperature-dependent exchange stiffness and domain wall width in Co” Phys. Rev. B **94**, 104433 (2016).
[36] L. Rózsa, U. Atxitia, and U. Nowak, “Temperature scaling of the Dzyaloshinsky-Moriya interaction in the spin wave spectrum” Phys. Rev. B **96**, 094436 (2017).
[37] Q. Gong, M. Yi, R.F.L. Evans, B.-X. Xu, and O. Gutfleisch, “Calculating temperature-dependent properties of Nd2Fe14B permanent magnets by atomistic spin model simulations” Phys. Rev. B **99**, 214409 (2019).
[38] J.D. Agudelo-Giraldo, H.H. Ortiz-Alvarez, J. Restrepo, and E. Restrepo-Parra, “Magnetic atomistic modelling and simulation of nanocrystalline thin films” Superlattices and Microstructures **105**, 90 (2017).
[39] J.D. Agudelo-Giraldo, E. Restrepo-Parra, and J. Restrepo, “Grain boundary anisotropy on nano-polycrystalline magnetic thin films” Scientific Reports **10**, 5041 (2020).
[40] D. Weller, G. Parker, O. Mosendz, E. Champion, B. Stipe, X. Wang, T. Klemmer, G. Ju, and A. Ajan, “A HAMR Media Technology Roadmap to an Areal Density of 4 Tb/in2” IEEE Trans. Magn. **50**, 3100108 (2014).
[41] I.M. Obaidat, B. Issa, and Y. Haik, “Magnetic Properties of Magnetic Nanoparticles for Efficient Hyperthermia” Nanomaterials **5**, 63 (2015).

[42] J. Waters, A. Berger, D. Kramer, H. Fangohr, and O. Hovorka, “Identification of Curie temperature distributions in magnetic particulate systems” J. Phys. D: Appl. Phys. **50**, 35LT01 (2017).